\documentclass[CCO,PDF,12pt]{cco_author} % use PDF command to enable PDFLaTeX driver
\usepackage{layout}
\usepackage{algorithm}
\usepackage[noend]{algpseudocode}
\usepackage{extraplaceins}
\usepackage{amsmath,amssymb}
\usepackage{amsfonts}

\title{An $O(n^3)$ time algorithm for the maximum-weight limited-capacity many-to-many matching}
\shorttitle{The maximum-weight limited-capacity many-to-many matching}
\author{Fatemeh~Rajabi-Alni\inst{1}\footnote{Corresponding Author},Behrouz~Minaei-Bidgoli\inst{1} 
       }

\shortauthor{F. Rajabi-Alni, B. Minaei-Bidgoli}

\institute{\inst{1}
           School of Computer Engineering, Iran University of Science and Technology, Tehran, Iran\\
           rajabi\_fatemeh@mail.iust.ac.ir, b\_minaei@iust.ac.ir
          }
\abstract{Given an undirected bipartite graph $G=(A \cup B, E)$, a \textit {many-to-many matching} (MM) in $G$ matches each vertex $v$ in $A$ (resp. $B$) to at least one vertex in $B$ (resp. $A$). In this paper, we consider the \textit {limited-capacity many-to-many matching} (LCMM) in $G$, where each vertex $v\in A\cup B$ is matched to at least one and at most $Cap(v)$ vertices; the function $Cap : A\cup B \rightarrow \mathbb{Z}> 0$ denotes the capacity of $v$ (an upper bound on its degree in the LCMM). We give an $O(n^3)$ time algorithm for finding a maximum (respectively minimum) weight LCMM in $G$ with non-positive real (respectively non-negative real) edge weights, where $\lvert A \rvert+\lvert B \rvert=n$.}

\history{}% will be inserted by the publisher
\begin{document}
\maketitle

\section{Introduction}
\label{intro}

Given two sets of objects $A$ and $B$, a \textit{many-to-many matching} (MM) matches each object of $A$ (respectively $B$) to at least one object of $B$ (respectively $A$). The MM has many uses, including computational biology, pattern recognition, and wireless networks \cite{LoC,Song,Rubert,Zhang}. We can represent the sets and their relations using a bipartite graph; for example, one part can represent mutated genes and the other part outlying genes \cite{Song}. Given an undirected bipartite graph $G=(A \cup B,E)$, a \textit{matching} in $G$ is a set of vertex disjoint edges $M \subseteq E$. Denote by $W(e)$ the weight of the edge $e\in E$. The weight of $M$ which is denoted by $W(M)$ is the sum of the weights of all the edges in $M$, hence $$W(M)=\sum_{e \in M}W(e).$$
A \textit {maximum weight matching} (MWM) denoted by $M'$ is a matching that for any other matching $M''$ we have $W(M'') \le W(M')$.
A \textit{perfect matching} is a subset of edges $PM \subseteq E$ such that every vertex of $G$ is adjacent to exactly one edge of $PM$. Assume that $\lvert A \rvert=\lvert B \rvert=n$ and $\lvert E\rvert=m$. The first polynomial-time algorithm for computing a \textit{maximum weight perfect bipartite matching} (MWPBM), a maximum weight perfect matching in the bipartite graph $G$, is the basic Hungarian algorithm \cite{Kuhn,Munkres}. Then, Fredman and Tarjan \cite{Fredman} solved the MWPBM problem in $G$ in $O(mn + n^2 \log n)$ time by implementing the Hungarian algorithm using fibonacci heaps. Later, other algorithms were developed for bipartite graphs with integer edge weights \cite{Gabow,Orlin}. For more discussion, see \cite{Imanparast}.
%his is often referred to as the minimum-weight edge cover problem.
%, running in $O(n^3)$ time

In many applications, the capacities of objects are limited. For example, consider a set of base stations communicating with a set of wireless sensors. The aim is to send data gathered by the sensors to the base stations. The number of sensors that can communicate with each base station is limited by the finite battery storage capacity of the sensors and the limited capacity of radio links of the base stations.
%Given an undirected bipartite graph $G=(A\cup B,E)$,

The \textit{capacity} of a vertex $v\in A\cup B$ denoted by a function $Cap : A\cup B \rightarrow \mathbb{Z} > 0$ is the number of vertices that can be matched to $v$. A \textit{limited-capacity MM} (LCMM) in $G$ is an edge set such that $1\leq deg(v)\leq Cap(v)$ for all $v \in A\cup B$ (we use $deg(v)$ to refer to the degree of the vertex $v$ in the LCMM). The maximum weight LCMM problem has been solved in the time complexity of $O(W'\sqrt{\beta'}m)$ for integer edge weights \cite{Huang}, where $W'=\max_{e \in E}(W(e))$ and $\beta'=\sum_{v \in A\cup B}Cap(v)$.

In this paper, we present an $O(n^3)$ time algorithm for the maximum weight LCMM problem in $G$ when the edge weights are non-positive real numbers. Note that a maximum weight LCMM in $G$ with non-positive real edge weights $W(e)$ is a minimum weight LCMM in $G$ with non-negative real edge weights $F(e)=-W(e)$ for all $e \in E$. An example of non-negative real edge weights is where the sets $A$ and $B$ are two sets of points in the plane (the weight of the edge between $p\in A$ and $q\in B$ equals the Euclidean distance between $p$ and $q$, thus $W(e)\geq 0$ for all $e\in E$).% An $O(n^2 poly (\log n))$ time algorithm have been proposed for the minimum cost many-to-many matching between two sets of points in the plane using the Hungarian algorithm. Note that the weight of the edge between $p$ and $q$ equals the Euclidean distance between them, thus $W(e)\geq 0$ for all $e\in E$. Therefore, our algorithm
%virThe

In a degree constrained subgraph (DCS) $H'=(V',E_{h'})$ of a general graph $H =(V',E_h)$ it holds that $l(v) \leq deg(v) \leq u(v)$ for each vertex $v \in V'$ with degree $deg(v)$, where $l(v)$ and $u(v)$ denote integer bounds. The maximum weight LCMM problem in $G$ can be stated as a specific case of the maximum weight DCS problem in a general graph and solved in $O(n^2 \min (m\log n,n^2))$ time \cite{Gabow1983}. We first review the basic Hungarian algorithm and some preliminary definitions. Then, we present our new algorithm.
%
%$H'=(V',E_{h'})$

\section{Preliminaries}

Let $G=(A \cup B,E)$ be a weighted bipartite graph such that $\lvert A \rvert=\lvert B \rvert=n$, $\lvert E \rvert=m$ and the edge weights are non-positive real values. Let $W(a,b)$ denote the weight of the edge $(a,b)$ for $a \in A$ and $b \in B$. If there exists no edge between two vertices $a_i\in A$ and $b_j \in B$, we assume that $W(a_i,b_j)=-\infty$. A path with the edges alternating between the edges of the matching $M$ and $E-M$ is called an \textit{alternating path}. Each vertex $v$ that is incident to an edge in $M$ is called a \textit{matched vertex}; otherwise, it is a \textit{free vertex}. All alternating paths originating from a free vertex
$v\in A\cup B$ constitute an \textit{alternating tree}. An alternating path with two free endpoints is called an \textit{augmenting path}. Note that by finding the symmetric difference between $M$ and an augmenting path, an \textit{augmentation} which is denoted by $augment(M)$, we get a new matching $M'$ with $\lvert M'\rvert =\lvert M\rvert+1$ (the size of $M$ increases by $1$).
%preedit: Note that if the edges of an augmenting path that are in $M$ are replaced with the ones that are in $E-M$, an \textit {augmentation}, its size increases by $1$.

A \textit{vertex labeling} is a function $l: V \rightarrow \mathbb{R}^- \cup \{0\}$ with $V=A \cup B$ that assigns a non-positive real value as a label to each vertex $v \in V$. A vertex labeling that in which $l(a)+l(b) \ge W(a,b)$ for all $a \in A$ and $b \in B$ is called a \textit {feasible labeling}. The \textit{equality graph} of a feasible labeling $l$ is a graph $G_l=(V,E_l)$ such that $E_l=\{(a,b)\in E\vert l(a)+l(b)=W(a,b)\}$. The set of the \textit{neighbors} of a vertex $u \in V$ is defined as $N_l(u)=\{v\in V\vert (v,u) \in E_l\}$. Consider a set of vertices $S \subseteq V$, we define $N_l(S)=\bigcup_{u \in S} N_l(u)$ as the set of the neighbors of $S$.

%The algorithm finds an augmenting path by building a tree, called the \textit{alternating tree}, via a breadth-first search, rooted at $v$ for all free vertices $v in A$.
%Starting from the free vertex $v$, the algorithm aims to find an augmenting path by building a tree, called the \textit{alternating tree}, via a breadth-first search, rooted at $v$.

%\newtheorem{lemma}{Lemma}
\begin{lemma}
\label{lem1}
Consider a feasible labeling $l$ of an undirected bipartite graph $G=(A\cup B, E)$. Let $S \subset A$ and $T\subset B$ with $N_l(S)=T$, where $S$ and $T$ represent the vertices of an alternating tree. Let
$$\alpha_l=\min_{a_i \in S, b_j \notin T}(l(a_i)+l(b_j)-W(a_i,b_j)).$$ If the labels of the vertices of $G$ are updated such that:
$$l'(v)=\left\{
\begin{array}{lr}
l(v)-\alpha_l & if \  v \in S
 \\
 l(v)+\alpha_l & if\  v  \in T
 \\
 l(v) & Otherwise
 \end{array}
\right.,$$
then $l'$ is a feasible labeling such that $E_l \subset E_{l'}$.
\end{lemma}

%\noindent \textbf{Proof.}
\begin{proof}
Obviously, in the cases $a \in S,b \in T$, or $a \notin S,b \notin T$, or $a \notin S,b \in T$, we have:
$$l'(a)+l'(b)\ge l(a)+l(b)\ge W(a,b).$$
And, for some vertices $a \in S,b \notin T$, we have
$$l'(a)+l'(b)=l(a)-\alpha_l+l(b)=W(a,b).$$
\end{proof}

%%%\qed

%\newtheorem{theorem}{Theorem}
\begin{theorem}
Let $l$ be a feasible labeling such that $E_l$ covers all vertices. If $M$ is a perfect matching in $E_l$, then $M$ is a maximum weight matching \cite{Kuhn}.
\end{theorem}

\begin{proof}
Suppose that $M'$ is a perfect matching in $G$, since each vertex is incident to exactly one edge of $M'$ we have:
$$W(M')=\sum_{(a,b) \in M'} W(a,b)\le \sum_{v \in (A \cup B)}l(v).$$ Thus, $\sum_{v \in (A \cup B)}l(v)$ is an upper bound for each perfect matching. Now assume that $M$ is a perfect matching in $E_l$:
$$W(M)=\sum_{e \in M}W(e)=\sum_{v \in (A \cup B)}l(v).$$
\end{proof}

%\qed

Now, we review the basic Hungarian algorithm which computes an MWPBM in an undirected bipartite graph $G=(A \cup B, E)$ with $\lvert A\rvert=\lvert B\rvert=n$ (see Algorithm \ref{BasicHungarian}). It has been shown that the maximum weight matching problem in bipartite graphs can be reduced to the MWPBM problem and solved using the Hungarian algorithm in $O(n^3)$ time \citep{Eiter}. Note that, for all the free vertices $v \in A$, the Hungarian algorithm builds an alternating tree rooted at $v$ to find an augmenting path.%does a breadth-first search and

%figure1
\begin{figure}[h!]
\vspace{-0.2cm}
\hspace{0cm}

\resizebox{1\textwidth}{!}{%
\includegraphics{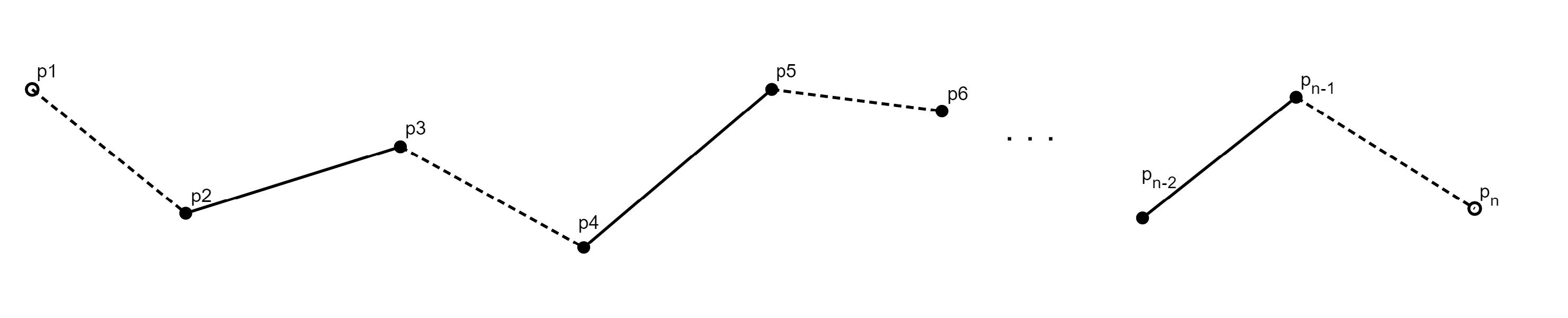}

}
%\vspace{5cm}       % Give the correct figure height in cm
\vspace{-0.5cm}
\caption{An example for illustration of Lemma \ref{maximum}; the hollow circles are free vertices and the filled circles are matched vertices.}
\label{fig:2}       % Give a unique label
\end{figure}

\algsetblock[Name]{Initial}{}{3}{1cm}
\alglanguage{pseudocode}
\begin{algorithm}

\caption{The Basic Hungarian algorithm($G=(A\cup B, E)$)}
\label{BasicHungarian}
\begin{algorithmic}[1]
\Initial \Comment Find an initial feasible labeling $l$ and a matching $M$ in $E_l$
\State $l(b_j)=0$ for all $1 \le j \le n$
\State $l(a_i)=\max_{j=1}^n W(a_i,b_j)$ for all $1 \le i \le n$
\State $M= \emptyset$
\While {$M$ is not perfect}

\State Select a free vertex $a_i  \in A$, and let $S = \{a_i\}$, $T=\emptyset$
\For{$j\gets 1$ to $n$}
\State $slack[j]=l(a_i)+l(b_j)-W(a_i,b_j)$
\EndFor
  \Repeat
     \If {$N_l(S)=T$}
        \State $\alpha_l=\min_{b_j \notin T}slack[j]$
        \State $Update(l)$ \Comment Update the labels according to Lemma \ref{lem1}
\ForAll {$b_j \notin T$}
 \State $slack[j]=slack[j]-\alpha_l$
\EndFor
        \EndIf
         \State Select $u  \in N_l (S)\setminus T$
          \If {$u$ is not free}\Comment ($u$ is matched to a vertex $z$; extend the alternating tree)
            \State $S = S \cup \{z\},T = T \cup \{u\}$.
\For{$j\gets 1, n$}
 \State $slack[j]=\min (l(z)+l(b_j)-W(z,b_j),slack[j])$
\EndFor

           \EndIf
          \Until {$u$ is free}
    \State Add $(a_i,u)$ to $M$ or $augment(M)$ so that all new adding edges to $M$ are in $E_l$
%$Augment(M)$Add $(a_i,u)$ to $M$ or
\EndWhile
\State \Return $M$
\end{algorithmic}
 \end{algorithm}

In Lines 2 and 3 of Algorithm \ref{BasicHungarian}, the vertices of the input bipartite graph are labeled by a feasible labeling. $M$ is an initial matching that can be empty (Line 4). In each iteration of the while loop of Lines 5--27, the size of $M$ increases by $1$, so it iterates at most $n$ times.% Let $$slack[j]=\min_{a_i\in S}(l(a_i)+l(b_j)-W(a_i,b_j)),$$
%by advantage of the array $slack[1,\dots,n]$, each iteration of the while loop can be run in $O(n^2)$ time.

The repeat loop of Lines 10--25 iterates at most $O(n)$ times until finding a free vertex $u$. In Line 12, the value of $\alpha_l$ is computed by:
$$\alpha_l=\min_{b_j \notin T}slack[j],$$
in $O(n)$ time. After computing $\alpha_l$, the feasible labeling $l$ is updated in Line 13 such that $N_l(S) \neq T$. The values of the slacks must also be updated by (Lines 14--15):
$$slack[j]=slack[j]-\alpha_l,$$ for all $b_j \notin T$. A vertex $u \in N_l(S) \setminus T$ is selected in Line 18. Observe that if $u$ is not a free vertex, the alternating tree should be extended (Lines 19--20). Note that in the repeat loop, an alternating tree is constructed to find an augmenting path. Once a vertex is moved from $\bar S$ to $S$, the values of $slack[j]$ for all $1\leq j\leq n$ are updated in $O(n)$ time (Lines 21--22). At most $O(n)$ vertices are moved from $\bar S$ to $S$, so the repeat loop takes the total time of $O(n^2)$. Therefore, the time complexity of the basic Hungarian algorithm is $O(n^3)$.

\begin{lemma}
\label{maximum}
After each augmentation of a matching (Line 26 of Algorithm \ref{BasicHungarian}), the cost of the matching does not increase.
\end{lemma}

\begin{proof}
Given an augmenting path $P$, two cases arise:

\begin{itemize}
\item $P=(p_1,p_2)$. According to non-positive real edge weights, this condition is trivial.

\item $P=(p_1,p_2,p_3,\dots,p_n)$ (see Figure~{\ref{fig:2}}). Note that $$W(p_i,p_{i+1})=l(p_i)+l(p_{i+1})$$ for $i=1,2,\dots,n-1$, since all the edges of an augmenting path are in $E_l$. Assume for a contradiction that the lemma is false, and thus

\begin{equation}
    \nonumber
    \begin{split}
    W(p_1,p_2)+W(p_3,p_4)+\dots+W(p_{n-1},p_{n})>
    &W(p_2,p_3)+W(p_4,p_5)\\
    &+\dots+W(p_{n-2},p_{n-1}).
\end{split}
\end{equation}

% \begin{eqnarray}
%  \nonumber W(p_1,p_2)+W(p_3,p_4)+\dots+W(p_{n-1},p_{n})>  & \\
% \nonumber W(p_2,p_3)+W(p_4,p_5)+\dots+W(p_{n-2},p_{n-1}).&
%\end{eqnarray}
    So, it holds that:
%    \begin{eqnarray}
%  \nonumber l(p_1)+l(p_2)+\dots+l(p_n)>  & \\
% \nonumber l(p_2)+l(p_3)+\dots+l(p_{n-1}),&
%\end{eqnarray}

\begin{equation}
    \nonumber
    \begin{split}
    l(p_1)+l(p_2)+\dots+l(p_n)>l(p_2)+l(p_3)+\dots+l(p_{n-1}),
\end{split}
\end{equation}

and thus: $$l(p_1)+l(p_n)>0.$$ Note that both $p_1$ and $p_n$ are free, and according to the above feasible labeling we have $l(p_n)\leq0$ and $l(p_1)\le 0$. Contradiction.

\end{itemize}\end{proof}

The Hungarian algorithm gives the best run time for bipartite graphs with low range edge weights \cite{LOPES201950}. In the worst case, the repeat loop of the algorithm runs in $O(n^2)$ overall time; the function $Update(l)$ of the algorithm produces a new feasible labeling $l'$ whose equality graph has only one more edge ($\lvert E_{l'}\rvert =\lvert E_l\rvert +1$). However, in bipartite graphs with low range edge weights and dense graphs, after updating the labels, many more edges are added to $E_l$, substantially decreasing the computational complexity. Low range matrices are used in problems with low precision data.

\section*{The maximum weight LCMM algorithm on bipartite graphs}
\label{newalgorithms}

Let $G=(A \cup B,E)$ be a bipartite graph with non-positive real edge weights, where $A=\{a_1,a_2,\dots,a_s\}$ and $B=\{b_1,b_2,\dots,b_t\}$ such that $s+t=n$. Let $C_A=\{\alpha_1,\alpha_2,\dots,\alpha_s\}$ and $C_B=\{\beta_1,\beta_2,\dots,\beta_t\}$ denote the capacities of $A$ and $B$, respectively. Assume w.l.o.g that $t\leq s$. Also, assume that $s\leq \sum_{j=1}^{t}\beta_j$ (it is obvious that if $s> \sum_{j=1}^{t}\beta_j$, then there does not exist any LCMM in $G$). We present an $O(n^3)$ time algorithm for computing a maximum weight LCMM in $G=(A \cup B,E)$, where each vertex $a_i \in A$ must be matched to at least one and at most ${\alpha }_i$ vertices in $B$, and each vertex $b_j \in B$ must be matched to at least one and at most ${\beta }_j$ vertices in $A$ for all $1 \leq i \leq s$ and $1 \leq j \leq t$.%($a_i$ and $b_j$ are connected by a infinite-weight edge)

\begin{figure}[h!]%[!tbp]
%\vspace{0cm}
\hspace{-0.2cm}

\resizebox{1.2\textwidth}{!}{%
\includegraphics{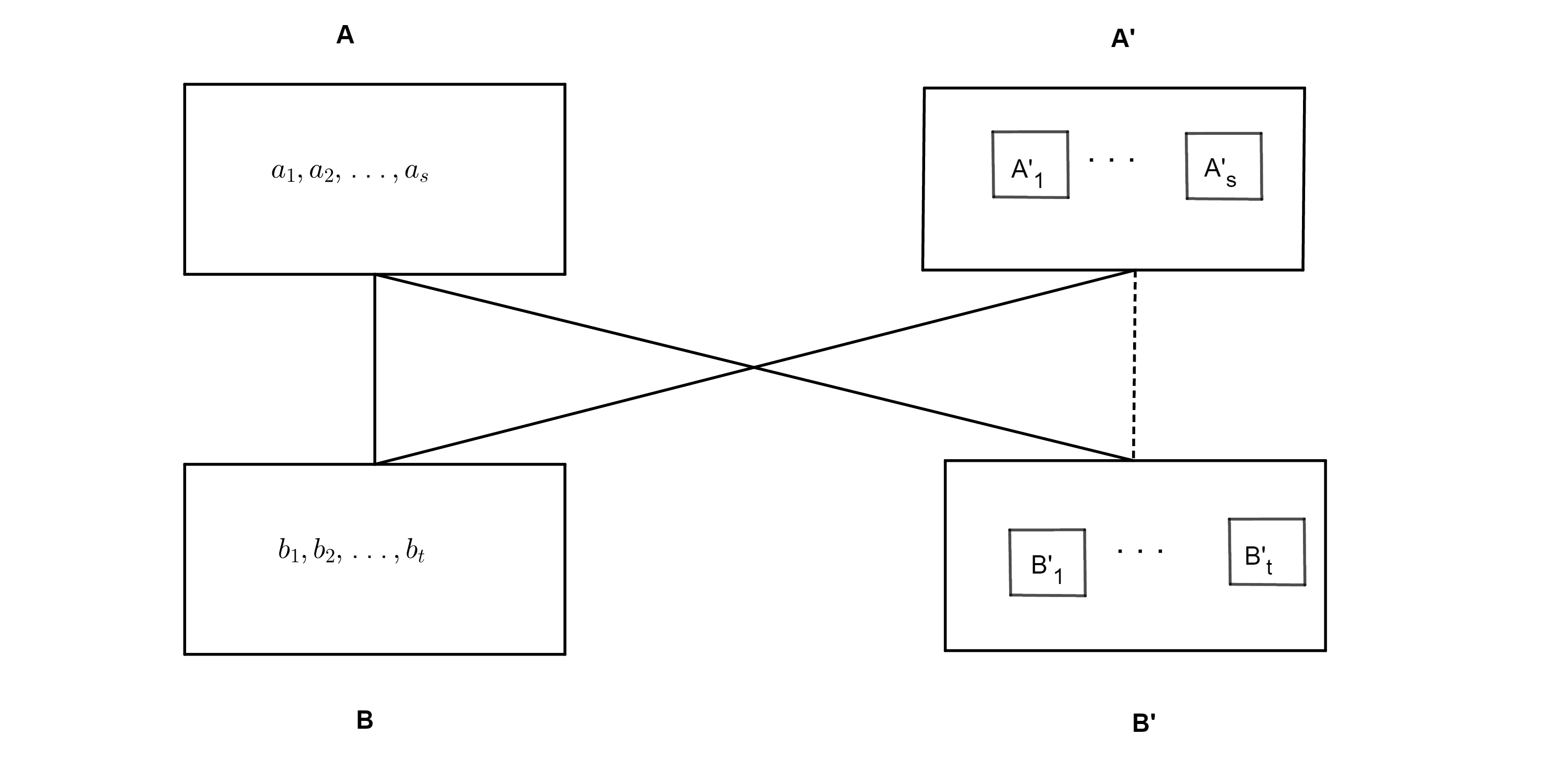}
}
\caption{Our constructed complete bipartite graph; the dashed line represents a complete connection with zero weighted edges between $A'$ and $B'$.}
%\vspace{5cm}       % Give the correct figure height in cm
%\vspace{-0.9cm}

\label{fig:1}       % Give a unique label
\end{figure}

Our algorithm is based on the basic Hungarian algorithm. Recall that the basic Hungarian algorithm computes an MWPBM in an undirected bipartite graph. Thus, firstly, we construct a complete bipartite graph $G'=(X \cup Y, E')$ with $X=A \cup A'$ and $Y=B \cup B'$ as follows (see Figure~{\ref{fig:1}}). Then, we run our algorithm on it.

In a \textit {complete connection} between two sets, each element of one set is connected to all the elements of the other set. We show each set of vertices by a rectangle and the complete connection between two sets by a line connecting the two corresponding rectangles.

Given $A= \{a_1, a_2, \dots, a_s\}$ and $B= \{b_1, b_2, \dots, b_t\}$, we construct a complete connection between $A$ and $B$, where the weight of the edge $(a_i, b_j)$ is equal to the cost of matching the element $a_i$ to the element $b_j$ for all $1 \le i \le s$ and $1 \le j \le t$.

Let $B'_j=\{b'_{j1},b'_{j2}, \dots, b'_{j(\beta_j-1)}\}$ for $1 \le j \le t$, and $B'= \{B'_1, B'_2, \dots, B'_t\}$. Each vertex of $A$ is connected to all the vertices of $B'$ such that $$W(a_i,b'_{jk})=W(a_i,b_j)$$ for all $1 \le k \le (\beta_j-1)$.

Let $A'_i=\{a'_{i1},a'_{i2}, \dots, a'_{i(\alpha_i-1)}\}$ for $1 \le i \le s$ and $A'= \{A'_1,A'_2, \dots, A'_s\}$, we also construct a complete connection between the sets $B$ and $A'$ such that $$W(a'_{ik},b_j)=W(a_i,b_j)$$ for all $1 \le k \le (\alpha_i-1)$. Also, there exists a complete connection between two sets $A'$ and $B'$ with zero weighted edges.

%shekl2

Now, we modify the basic Hungarian algorithm to get a new algorithm, called $ModifiedHungarianAlg$ (see Algorithm \ref{ModifiedHungarian}). In the modified Hungarian algorithm, Line 5 of Algorithm \ref{BasicHungarian} is changed; the while loop iterates until matching the subset $A\subseteq X$. The initialization step is also removed.

Our new algorithm consists of two steps (see Algorithm \ref{CMWBM}): in the first step, the vertices of $A\subseteq X$ are matched, and in the second one, the vertices of $B\subseteq Y$. We claim that by applying our algorithm on $G$, $Maxweight\  LCMM \ Algorithm(G=(A \cup B,E))$, we get a maximum weight LCMM in $G$.

\begin{enumerate}
\item[Step I.]Given an undirected complete bipartite graph $G'=(X\cup Y,E')$ with $X=A \cup A'$ and $Y=B \cup B'$, in this step, we call the function $ModifiedHungarianAlg(G',A,M,l)$ (Line 6 of Algorithm \ref{CMWBM}). It matches the vertices $a_i \in A$ for all $i=\{1,2,\dots,s\}$ until there does not exist any free vertex in $A\subseteq X$.

    The while loop of Lines 1--30 of $ModifiedHungarianAlg(G',A,M,l)$, called \textit{the main loop}, iterates until all the vertices of $A$ are matched to exactly one vertex of $B \cup B'$. Obviously, it iterates $O(n)$ times, since the number of vertices of $A$ is $O(n)$. In the following, we show that each iteration of the main loop of $ModifiedHungarianAlg(G',A,M,l)$ takes $O(n^2)$ time.

\newtheorem{Observation}{Observation}
\begin{Observation}
\label{labelsB'}
The labels of all the free vertices $b'_{jk}\in B'_j$ are equal for all $1 \le k \le \beta_j-1$.
\end{Observation}

Initially, we have $l(b'_{jk})=0$ for all $1 \leq k \leq \beta_j-1$. The function $Update(l)$ updates the labels of all the vertices $b'_{jk}\in T$, i.e., all the vertices $b'_{jk}$ that have been matched to a vertex in $S$. Hence, all the free vertices $b'_{jk}\in B'_j$ have equal labels for $1 \le k \le \beta_j-1$.

\begin{Observation}
\label{slackB'}
The slacks of all the free vertices $b'_{jk}\in B'_j$ are equal for all $1 \le k \le \beta_j-1$.
\end{Observation}

%Note that
%Observation \ref{labelsB'} implies that the values of the slacks of all free vertices $b'_{jk}\in B'_j$ with $1 \le k \le \beta_j-1$ are equal.

Note that the vertices of $B'_j$ are copies of the vertex $b_j$. By Observations \ref{labelsB'} and \ref{slackB'}, all the free vertices of $B'_j$ have equal labels and slacks, so in each iteration of the main loop, we consider only one of the free vertices $b'_{jk}\in B'_j$ for $1 \le k \le \beta_j-1$, arbitrarily. Actually, in each iteration of the main loop, all the free vertices $b'_{jk}\in B'_j$ are considered as a single vertex (Line 5 of Algorithm \ref{ModifiedHungarian}).

Let $B''={b''_1,b''_2, \dots,b''_t}$, where $b''_j$ is an arbitrary free vertex of $B'_j$, if exists (Line 6 of Algorithm \ref{ModifiedHungarian}). Let $Y'=B \cup C \cup B''$, where $C$ is the set of the matched vertices of $B'$ with respect to $M$ (Lines 8--9 of Algorithm \ref{ModifiedHungarian}). In each iteration of the main loop, we first give the slacks of all the vertices $y_j \in Y'$ initial values in $O(n)$ time (Lines 10--12). Then, the repeat loop of Lines 13--28 iterates until we find a free vertex $u$ and add $(x_i,u)$ to $M$ or $augment(M)$. Note that if $N_l(S)=T$, we update the labels of the vertices of $G'$ to get a new feasible labeling such that $N_l(S)\neq T$. The neighbor set of a vertex $u \in A$ is defined as $N_l(u)=\{v \in Y'\vert (u,v) \in E_l\}$.

Note that there exist at most $O(n)$ matched vertices, i.e. the vertices of $A$, so the numbers of the vertices of $T$ and $S$ are at most $O(n)$. Hence, in Line 15, we get the minimum value in $O(n)$ time. Also, updating the labels and slacks takes $O(n)$ time (Lines 16, 17--19 and 24--26).

 Observe that in this step, the initial matching $M$ is an empty set (Line 5 of Algorithm \ref{CMWBM}), and each iteration of the main loop starts from an arbitrary free vertex $a_i \in A$ for $1\leq i\leq s$. Therefore, by Lemma \ref{maximum}, the output of this step, $M_1$, is an MWM covering $A$. Note that in $M_1$, all the vertices of $A$ are matched, but some vertices of $B$ might be free.

\item[Step II.] In this step, we call the function $ModifiedHungarianAlg(G',B,M_1,l_1)$ (Line 7 of Algorithm \ref{CMWBM}). We use the final labels and slacks of the vertices from Step I, so the labels of the vertices are feasible. We also use the output matching of Step I, $M_1$, as the initial matching of Step II. Once all the vertices of $B$ are matched, this step terminates. Recall that the initial matching of this step, $M_1$, is an MWM that covers $A$. Additionally, observe that each iteration of the main loop of $ModifiedHungarianAlg(G',B,M_1,l_1)$ starts from an arbitrary free vertex $b_j \in B$ for $1\leq j\leq t$. Therefore, the output of this step, $M_2$, is an MWM covering $A\cup B$. Note that, by Lemma \ref{maximum}, if we continue matching the vertices of $G'$, the cost of the matching does not increase. Similar to Step I, we can show that the time complexity of Step II is also $O(n^3)$. We observe that the slacks and labels of all the free vertices $a'_{ik}\in A'_i$ for all $1 \le k \le \alpha_i-1$ are equal in value.
\end{enumerate}

We claim that from the output matching of Step II, $M_2$, we get a maximum weight LCMM in $G=(A \cup B,E)$, denoted by $L$. In the following, we prove that the weight of $M_2$ is equal to the weight of $L$, i.e., $W(L)=W(M_2)$.

%\begin{theorem}
%\label{new2}
%$W(L)= W(M_2)$.
%\end{theorem}

%\begin{proof}

\begin{lemma}
\label{l1}
$W(L)\leq W(M_2)$.
\end{lemma}

\begin{proof}
We get from $L$ a matching $M'$ in $G'$ covering $A\cup B$ such that $W(M')=W(L)$. We relate an edge of $G'$ to each pairing of $L$ as follows.

  For each $(a_i,b_j)\in L$, three cases arise:
  \begin{itemize}
    \item Both $a_i\in G'$ and $b_j\in G'$ are free. Then, we add $(a_i,b_j)$ to $M'$.
    \item $a_i\in G'$ is free but $b_j\in G'$ is matched. Then, we add $(a_i,b'_{jk})$ to $M'$, where $b'_{jk}$ is an arbitrary free vertex of $B'_j\in G'$.
    \item $a_i\in G'$ is matched but $b_j\in G'$ is free. In this situation, we add $(a'_{ik},b_j)$ to $M'$, where $a'_{ik}$ is an arbitrary free vertex of $A'_i\in G'$.
  \end{itemize}

\begin{figure}[h!]%[!tbp]
%\vspace{0cm}
%\hspace{0.7cm}

\resizebox{1.2\textwidth}{!}{%
\includegraphics{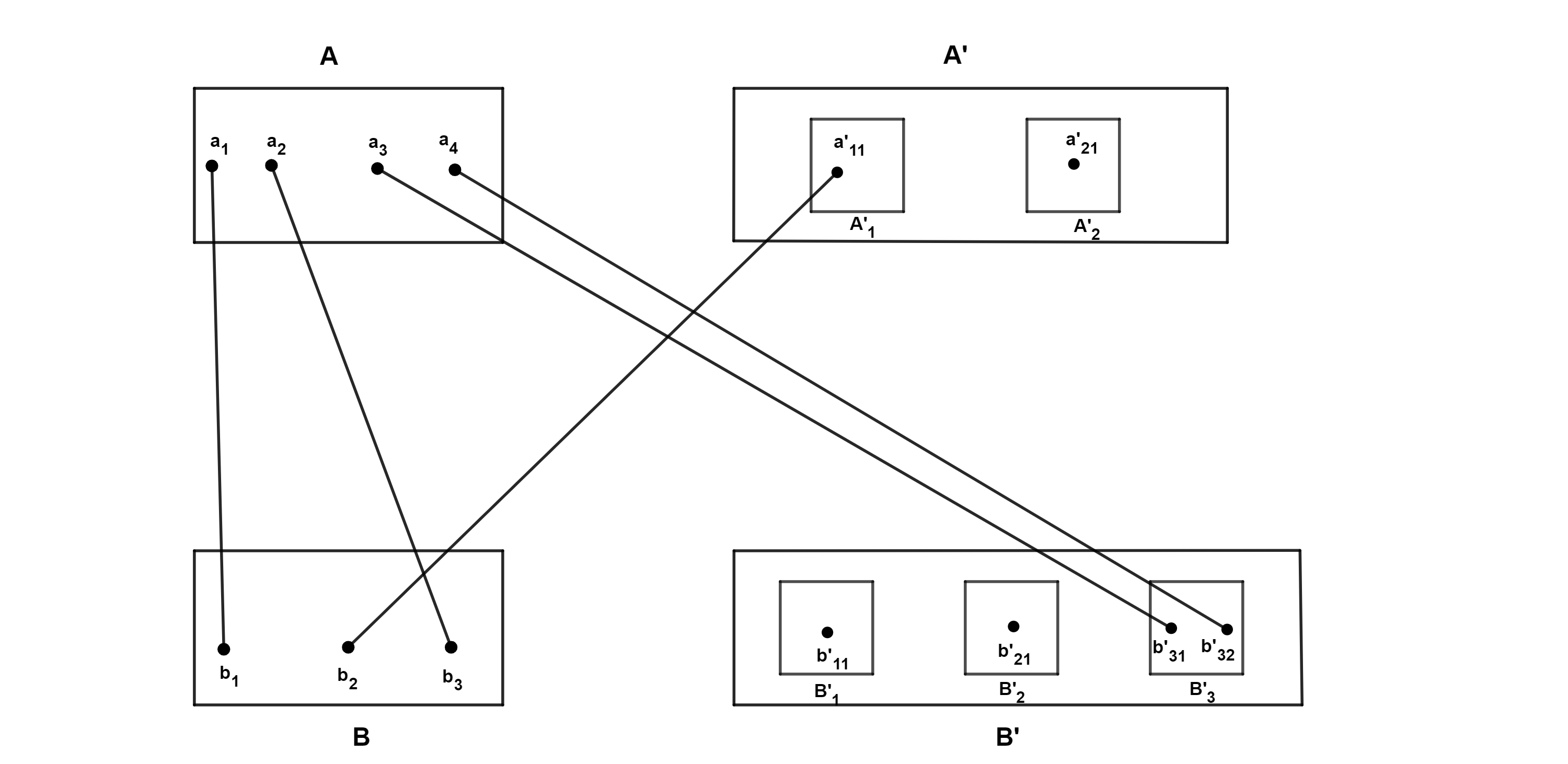}
}
\caption{The edges of $G'$ related to the pairings of $L=\{(a_1,b_1),(a_1,b_2),(a_2,b_3),(a_3,b_3),(a_4,b_3)\}.$}
%\vspace{5cm}       % Give the correct figure height in cm
%\vspace{-0.9cm}

\label{fig:new}       % Give a unique label
\end{figure}

%Recall that for all the edges $(p,q)$ with $p\in A'$ and $q\in B'$, we have $W(p,q)=0$. Thus, we can assume that there exist other edges $(p,q)\in M'$ with $p\in A'$ and $q\in B'$.
Thus, for each pairing $(a_i,b_j)\in L$ with no equivalent edge in $M'$, we add the edge $(a_i,b_j)$ to $M'$ if neither $a_i\in G'$ nor $b_j\in G'$ are incident with an edge in $M'$. Otherwise, we add to $M'$ the edge incident to the free vertex of $(a_i,b_j)$ and an arbitrary free vertex (see Figure \ref{fig:new} for an example).%(i.e., either $a_i\in G'$ or $b_j\in G'$)

For each $(a_i,b_j)\in L $, we add an edge with equal weight in $M'$, so $W(M')=W(L)$. $M_2$ is a maximum weight matching in $G'$  which covers $A\cup B$, that is $W(M')\leq W(M_2)$, so $W(L)\leq W(M_2)$.

\end{proof}

%Thus, we relate the edge $(a_i,b_j)\in G'$ to each pairing $(a_i,b_j)\in L$, if $a_i,b_j \in G'$ are free. Otherwise, we add if $a_i\in G'$ is free andwe add $(a_i,b'_{jk})$ to $M'$

In the following, we get an LCMM $L'$ in $G$ from the output matching $M_2$ in $G'$. Note that for each $a_i\in A$ for $1\leq i\leq s$, there exists the set $\{a_i\} \cup A'_i$ in $G'$ with $\alpha_i$ vertices. Also, there exist $\beta_j$ copies of each $b_j\in B$ in $G'$ (i.e., $\{b_j\} \cup B'_j$) for $1\leq j\leq t$. Therefore, the capacities of the vertices of $A$ and $B$ are satisfied in $M_2$.
For each edge $e \in M_2$, if $e=(a_i,b'_{jk} )$ for $1\leq k\leq \beta_j-1$, or $e=({a'}_{ik},b_j )$ for $1\leq k\leq \alpha_i-1$, or $e=(a_i,b_j)$, we add the pairing $(a_i,b_j )$ to $L'$. It is easy to see that $W(L)\geq W(L')=W(M_2)$.

% Note that for each edge of $M_2$, we add an edge with equal weight to $L$, thus $W(L)=W(M_2)$.
%\end{proof}

\begin{theorem}
Let $G=(A \cup B,E)$ be a non-positive real weighted bipartite graph with $\lvert A \rvert+\lvert B \rvert=n$, a maximum weight LCMM in $G$ can be computed in $O(n^3)$ time.
\end{theorem}

\algsetblock[Name]{Initial}{}{3}{1cm}
\alglanguage{pseudocode}
\begin{algorithm}

\caption{ModifiedHungarianAlg($G'=(X\cup Y,E')$,$A$,$M$,$l$)}
\label{ModifiedHungarian}
\begin{algorithmic}[1]
\While{$\{ u \in A\vert u \ is \ free\}\neq \emptyset$}

\State Select a free vertex $x_i  \in A$, and let $S = \{x_i\}$, $T=\emptyset$
\State $B''=\emptyset$
\For{$j\gets 1$ to $\lvert B\rvert $}
\State Select a free vertex $v  \in B'_j$
\State $B''=B'' \cup \{v\}$
\EndFor
\State Let $C$ be the set of the matched vertices of $B'$
\State $Y'=B \cup C \cup B''$

\ForAll {$y_j \in Y'$}
\State $slack[j]=l(x_i)+l(y_j)-W(x_i,y_j)$
\EndFor
  \Repeat
     \If {$N_l(S)=T$} \Comment $N_l(u)=\{v \in Y'\vert (u,v)\in E_l\}$
        \State $\alpha_l=\min_{y_j \in Y' \setminus T}slack[j]$
        \State $Update(l)$ \Comment Update the labels according to Lemma \ref{lem1}
\ForAll {$y_j \in Y' \setminus T$}
 \State $slack[j]=slack[j]-\alpha_l$
\EndFor
        \EndIf
         \State Select $u  \in N_l (S) \setminus T$
          \If {$u$ is not free}\Comment ($u$ is matched to a vertex $z$, extend the alternating tree)
            \State $S = S \cup \{z\},T = T \cup \{u\}$.
\ForAll {$y_j \in Y'$}
 \State $slack[j]=\min (l(z)+l(y_j)-W(z,y_j),slack[j])$
\EndFor

           \EndIf
          \Until {$u$ is free}
    \State Add $(x_i,u)$ to $M$ or $augment(M)$ so that all new adding edges to $M$ are in $E_l$
%$Augment(M)$
\EndWhile
\State \Return $M$ and $l$
\end{algorithmic}
 \end{algorithm}

\algsetblock[Name]{Initialize}{}{4}{1cm}

\alglanguage{pseudocode}
%\vspace{1cm}
\begin{algorithm}
\caption{Maxweight LCMM Algorithm($G=(A \cup B,E)$)}
\label{CMWBM}
\begin{algorithmic}[1]
\State Construct the bipartite graph $G'=(X\cup Y,E')$ from $G$ with $X=A \cup A'$ and $Y=B \cup B'$
%\State Find an initial feasible labeling $l$ and an initial matching $M$ in $E_l$
\Initial \Comment Find an initial feasible labeling $l$ and a matching $M$ in $E_l$
\State $p=\lvert X\rvert $, $q=\lvert Y\rvert $
\State Let $l(y_j)=0$ for all $1 \le j \le q$ and $l(x_i)=\max_{j=1}^q W(x_i,y_j)$ for all $1 \le i \le p$
\State $M=\emptyset$
\State $(M_1,l_1)=$\Call{ModifiedHungarianAlg}{$G'$,$A$,$M$,$l$}
\State $(M_2,l_2)=$\Call{ModifiedHungarianAlg}{$G'$,$B$,$M_1$,$l_1$}
\State \Return $M_2$
 \end{algorithmic}
\end{algorithm}

\section*{Conflict of Interest}

The authors declare that they have no conflict of interest.
\section*{Data Availability}

Data sharing is not applicable to this article as no datasets were
generated or analyzed during the current study.

\FloatBarrier

%\section*{Conclusions}
%In this paper, we proposed an $O(n^3)$ time algorithm for finding a maximum weight Cb-matching in a bipartite graph $G=(A\cup B,E)$ with $|A|+|B|=n$. We modified the basic Hungarian algorithm and presented an algorithm for a more general version of the matching problem. In the future, we hope to develop new algorithms for other versions of the b-matching problem depending on the properties of two given input sets.%; for example when there are a given number of duplicates in two given genomes.

%\bibliographystyle{sn-basic}
%\bibliographystyle{sn-vancouver}
%\bibliography{mybibbb59f2}

\end{document}